\documentclass[10pt,a4paper]{iopart}\usepackage{iopams}      

\usepackage{graphicx}

\usepackage{eucal} %jxzj
\usepackage{calrsfs} %jxzj
\usepackage{amssymb}%nova org
\usepackage{amsfonts}%nova org
\usepackage{amsbsy}%jxzj uncommond
\usepackage{mathrsfs}

\begin{document}

\title[]{Quantum Mechanical Probability  of Internally Electrodynamic Particle(s)}

\author{J.X. Zheng-Johansson
}
\address{
Institute of Fundamental Physics Research, 611 93 Nyk\"oping, Sweden  
} 
%\date{April, 2007}
%\date{December 17, 2006}
%\address{July, 2007}
%\address{November 7, 2007}
%\address{
%Dec 10,  2007; -;
%Nov 17, 2009; April 23-26,July 8th, Aug 05, Aug. 15,22,27,30, Sept 9, 14, 17,21b,Oct 14, 2010(Oct. 25, 2010)
%(submitted to group 28: Oct. 30, 2010)
%}

\ead{jxzj@iofpr.org}
%\maketitle

\def\vac{{\rm{vac}}}
\def\Vcal{{\mathscr{V}}} %use with \usepackage{calrsfs} %jxzj?\usepackage{mathrsfs}
\def\mini{0}
\def\Pcal{{\mathcal{P}}}
\def\bav{{\bar{b}}}
\def\v{{\rm v}}
\def\Hbar{\bar{H}}
\def\pbar{\bar{p}}

\def\D{\Delta}
\def\bcal{b}
\def\bbar{\mathbin{{b}\mkern-11.mu^{_{\mbox{\small{$-$}}}}\hspace{-0.08cm} }}
\def\nstat{\nu}
\def\nst{\nu}
\def\engbar{\bar{\eng}}
\def\engobar{\bar{\eng}_0}
\def\psias{\psi}
\def\Phimas{\Phim}
\def\fas{f}
\def\rbb{\as}

\def\La{L}
\def\Ja{J}
\def\as{p}
\def\ioii{{\mbox{\normalsize${\frac{1}{2}}$}}}
\def\Rb{{\bf R}}
\def\rb{{\bf r}}
\def\ub{{\bf{u}}}
\def\hatu{\hat{u}_q}
\def\Nsub{{{\mbox{\tiny${N}$}}}}
\def\Hsub{{{\mbox{\tiny${H}$}}}}

\def\Pisub{{{\mbox{\tiny${\mit{\Pi}}$}}}}

\def\q{\bar{q}}
\def\Nst{\Ncal}
\def\Nstat{\Ncal}
\def\xdot{\dot{x}}
\def\exc{{\rm exc}}
\def\ens{{\rm ens}}
\def\Lcal{\mathcal{L}}
\def\Tcal{\mathcal{T}}
\def\Kcal{{\mathcal{K}}}
\def\Xcal{{\mathcal{X}}}

\def\Wvel{\Omegavel}
\def\Ncal{\aleph}
\def\Omegavel{\mathbin{{\mit\Omega}\mkern-13.mu^{_{\mbox{$-$}}}\hspace{-0.08cm}{}_d }}
\def\omegavel{\mathbin{{\mit\omega}\mkern-13.mu^{_{\mbox{$-$}}}\hspace{-0.08cm}{}_d }}
\def\wvel{\omegavel}

\def\q{\mathbin{q\mkern-11mu-}}
\def\PE{\mbox{\tiny{{\rm P.E.}}}}
\def\ME{\mbox{\tiny{{\rm M.E.}}}}
\def\QM{\mbox{\tiny{{\rm QM}}}}
\def\Psub{\mbox{\tiny{{\rm P}}}}
\def\Bsub{{\mbox{\tiny{{\rm B}}}}}
\def\Asub{{\mbox{\tiny{{\rm A}}}}}

\def\ev{\epsilon}
\def\Ucal{\bar{\eng}_{0}}
\def\Omegavel{\mathbin{{\mit\Omega}\mkern-13.mu^{_{\mbox{$-$}}}\hspace{-0.08cm}{}_d }}
\def\omegavel{\mathbin{{\mit\omega}\mkern-13.mu^{_{\mbox{$-$}}}\hspace{-0.08cm}{}_d }}
\def\Wvel{\Omegavel}

\def\Ci{1}
\def\betamt{{\bf{b}}}
\def\kb{{\bar{k}}}
\def\kbf{{\bf{k}}}
\def\Kb{{\bf{K}}}
\def\cb{{\bf{c}}}

\def\pb{{\bar{p}}}
\def\pbf{{\bf{p}}}
\def\Acal{{\cal{A}}}
\def\Bcal{{I_{{\rm{ex}}}}}
\def\Ccal{{\cal{C}}}
\def\Vp{V}
\def\m{{{}_{\mbox{-}}}}
\def\Ccal{{\cal{C}}}
\def\p{{{}_{+\hspace{-0.1cm}}}}

\def\psipi{\psi_{\p}(1)}
\def\psipii{\psi_{\p}(2)}
\def\psimi{\psi_{\m}(1)}
\def\psimii{\psi_{\m}(2)}

\def\ai{\alpha(1)}
\def\aii{\alpha^{'}(2)}
\def\bi{\beta^{'}(1)}
\def\bii{\beta(2)}

\def\fa{f_r}
\def\fb{f_\ell}

\def\Ca{C_a}
\def\Cb{C_b}
\def\fbf{{\bf{f}}}
\def\Ocal{{\cal{O}}}
\def\psib{{\pmb{\psi}}}
\def\alphab{{\pmb{\alpha}}}
\def\sigmab{{\pmb{\sigma}}}

\def\Eb{{\bf E}}
\def\Bb{{\bf B}}
\def\ke{\kappa}
\def\nabb{{\pmb{\nabla}}}
\def\nablab{{\pmb{\nabla}}}
\def\vir{{\rm vir}}
\def\psitot{\psi}
\def\jb{{\bf{j}}}
\def\vel{v}
\def\velb{{\bf{v}}}

\def\Imtr{I}
\def\citeUnif{5}
\def\App{}
\def\Qcal{{\mathcal{Q}}}
\def\Tcal{{\mathcal{T}}}
\def\Cross{Q}

\def\vphilim{f}
\def\ft{{\mathcal{B}}}
\def\vphibar{\mathbin{\varphi\mkern-12.5mu-}}
\def\vphi{\varphi}
\def\med{{\med}}
\def\Mcal{{\mathfrak{M}}}
\def\Sb{{\bf{S}}}
         \def\xia{{\mathcal{A}}}
\def\tha{\theta}

\def\nb{\bf{n}}
\def\zb{{\bf{z}}^0}
\def\phiv{\varphi}
\def\Lb{{\bf{L}}}
\def\velsub{_{\vel}}

\def\nablab{{\pmb{\nabla}}}
\def\velb{{\pmb{\vel}}}
\def\minus{\mbox{-}}

\def\Ab{{\bf{A}}_a}
\def\vel{\upsilon}
\def\Thm{\vartheta}
\def\Thetam{{\mit{\Theta}}}
\def\lb{{\bf l}}
\def\vb{{\bf{v}}}

\def\Rb{{\bf R}}
\def\pd{\partial}
\def\vphi{\varphi}

\def\psitot{\varphi}
\def\psiR{\widetilde{\psi}}
\def\psiL{\widetilde{\psi}^{{\rm vir}}}
\def\Phim{{\mit{\Phi}}}
\def\PhimR{\widetilde{ {\mit \Phi}}}
\def\PsimR{\widetilde{ {\mit \Psi}}}
\def\PsimL{{\widetilde{ {\mit \Psi}}}^{{\rm vir}}}
\def\a{\alpha}
\def\uav{\bar{u}}
\def\D{\Delta}
\def\th{\theta}
\def\r{{\mbox{\tiny${R}$}}}
\def\re{{\mbox{\tiny${R}$}}}
\def\Fmed{F_{{\rm a.med}}}
\def\med{{\rm med}}
\def\Lw{L_{\varphi}}
\def\Fb{{\bf{F}}}

\def\Efb{{\bf{E}}}
\def\Bfb{{\bf{B}}}
\def\Ac{ \varphi}
\def\Xsub{{\mbox{\tiny${X}$}}}
\def\Ysub{{\mbox{\tiny${Y}$}}}
\def\Zsub{{\mbox{\tiny${Z}$}}}

\def\Ksub{{\mbox{\tiny${K}$}}}
\def\W{{\mit \Omega}}
\def\Wd{\W_d{}}
\def\Nu{{\cal V}}
\def\Nud{\Nu_d{}}
\def\Eng{{\cal E}}
\def\eng{{\varepsilon}}
\def\Acuni{\Ac_{{\Ksub}^\dagsup}^{\dagsup}}
\def\unduni{\Ac_{{\Ksub}^\dagger}^{\dagsup}}
\def\Acauni{\Ac_{{\Ksub}^\ddagsup}^{\ddagsup}}
\def\Acunim{{\Ac_{{\Ksub}^\dagsup}^{\dagsup *}}}
\def\undunim{{\Ac_{{\Ksub}^\dagsup}^{\dagsup}}^*}
\def\Acaunim{{\Ac_{{\Ksub}^\ddagsup}^{\ddagsup *}}}
\def\pd{\partial}
\def\Ad{ {\mit \psi}}
\def\psim{ {\mit \psi}}
\def\Kd{K_d{}}
\def\Lam{{\mit \Lambda}}
\def\lam{\lambda}
\def\dagsup{{\mbox{\tiny${\dagger}$}}}
\def\ddagsup{{\mbox{\tiny${\ddagger}$}}}
\def\psimKdK{\psim_{\Ksub,\Kdsub}}
\def\w{\omega{}}
\def\wdlow{\omega_d }
\def\g{\gamma{}} 
\def\Phimc{{\mathcal C}}
\def\Psim{{\mit \Psi}}
\def\arm{{\rm a}}
\def\brm{{\rm b}}
\def\crm{{\rm c}}
\def\drm{{\rm d}}
\def\erm{{\rm e}}
\def\frm{{\rm f}}
\def\grm{{\rm g}}
\def\hrm{{\rm h}}
\def\lf{\left}
\def\rt{\right}
\def\Kdsub{{\mbox{\tiny${K_d}$}}}
\def\psimkd{\psim_{\kdsub}}
\def\psimKd{\psim_{\Kdsub}}
\def\hquad{ \ \ } 
\def\Taum{{\mit \Gamma}}

\def\dagsup{{\mbox{\tiny${\dagger}$}}}
\def\ddagsup{{\mbox{\tiny${\ddagger}$}}}

%\tableofcontents

\begin{abstract}

A distribution of electromagnetic fields presents a statistical assembly of a particular type, which is at scale $h$ a quantum statistical assembly itself and has also  been instrumental to a concrete demonstration of the  basic probability assumption of quantum mechanics. Of specific concern in this discussion is an extensive wave train of radiation fields, described by a total wave function $\psi$, which are continuously (re)emitted and (re)absorbed  by an oscillatory (point) charge of a zero rest mass and yet a finite dynamical mass, with the waves and charge together making up an extensive undulatory IED particle. The IED particle will as any real particle be subject to interactions with the environmental fields and particles, hence to excitations, and therefore will explore all possible states over time; at scale $h$ the states are discrete. On the basis of the principles of statistics and statistical mechanics combined with first principles solutions for the IED particle, we derive for the specific statistical system of the IED particle the probability functions in position space, of a form $|\psi|^2$, and in dynamical-variable space.
%\\ \\
%*Presentation at the 28th Int Colloq Group Theo Meth in Phys, Univ Northumbria, UK, July 25-30, 2010; full paper submitted Oct 30, 2010.

%\draft
%\keywords{Probability, Quantum mechanical probability, Electric charge radiation,  Electromagnetic waves, Maxwell's equations, Classical wave equation, Quantum mechanical wave equation, Matter wave, Particle model, Unification scheme}

\end{abstract}

\section{Introduction}
It has been a  long established fact in experiment and formal theory  that  matter particles manifest both the characteristics of  corpuscle and wave, termed as matter waves ($\Psim$), and are at the scale of Planck constant $h$ dominated by quantum mechanics which formally is formulated on the basis of the wave equations of E Schr\"odinger, W Heisenberg and  P Dirac\cite{Merzbacher:1970}. 
The interpretation of the quantum mechanical wave function $\Psim$ is based on  the probability assumption of M Born\cite{Born1926}  that  the state of a  particle, assumed  corpuscular,  is completely specified by its $\Psim(x,t)$, with $\Psim(x,t)$ being generally complex, and $|\Psim(x,t)|^2$ represents the statistical probability of finding the particle at position $x$. 
This probability has a key   feature distinct from that of  classical systems:  it is additionally constrained by the Heisenberg uncertainty relations for the conjugate dynamical variables \cite{Heisenberg1930}. In phase space, this probability is associated with a smallest volume,  equal to the fundamental constant $h$ of M  Planck   raised to power $fN$ for a $N$ particle system of $f$ degrees of freedom, which is accessible to a microscopic state.
%for excitations. 

Quantum mechanics  has proven successful in predicting a broad range of quantum mechanical properties of particles, especially properties involving the quantization of dynamical variables, for which  the quantum theoretical predictions are  directly comparable with the overall vast quantity of  experimental data accumulated over a century or so to this date. 
However, it is up to the present not satisfactorily understood  that what is waving with a matter wave, what is accordingly  statistical with it in position as well as dynamical variable spaces, and what is the origin of  the Planck constant which defines a smallest volume in phase space? 

Their understandings come especially in question when our purpose is in order that the probability, and  the fundamental constant $h$ here,  
             %as the  specific concerns in this and a subsequent  paper, 
           %and a ramge of other fundamental physical properties and relations  
can be inferred based on a common minimal set of first principles' laws which also govern the remaining basic properties of particles. Further to this, quantum mechanics as interpreted through a statistical corpuscular-particle picture meets difficulties in (coherently) accounting for certain basic experimental phenomena, including in particular the basic coherent wave phenomena such as diffraction and (self) interference\cite{jxzjied}. Furthermore, the corpuscular particle with a given rest mass does not furnish a mechanical scheme for accounting for the origin of mass, the cause of gravity and the origin of relativistic effect, among others  (see \cite{jxzjied} for a recent fuller review).

On the whole, we appear to  have in hand a comprehensive, rigorous mathematical structure for quantum mechanics and yet an incomplete conceptual grasp of the underlying mechanical content. Conceptually, we are confronted with various internal clashes.   These concern  for examples the dual corpuscle and wave aspects in a single object, the desired wave form of spreading and the corpuscular particle assumption at any one time, and the relationships between prediction,  measurement  and actuality of the state of a particle as a subject of debate ever since the foundation of quantum mechanics. 
The difficulties and clashes, as one would identify upon an examination, are the inevitable results of descriptions 
based on  the corpuscular particle picture.
 Although, it can not be overemphasised that,  the current description represents the best theory which is as accurately as can be made in the given framework.
 
From the standpoint  that nature constructs and operates itself by mechanical means, indefinitely down to as small a scale as it takes to sum up to the lager-scale manifestations in question, one may anticipate, as it has already been demonstrated to certain extent in [\citeUnif a-m], that the conceptual  clashes will  all of a sudden  vanish once a realistic model of particle, one with a realistic internal structure and dynamical scheme, is constructed. 

Based on overall experimental observations as input information, the experimentally established de Broglie wave and relations from the outset, the author developed in [\citeUnif a-m] an internally electrodynamic  (IED) particle model,  which briefly states that  {\it a single-charged material particle, like the electron, proton, etc., is composed  of (i) an oscillatory point-like  charge $q$ (as source) with a characteristic frequency  $\W$ and zero rest mass,  and (ii) the  electromagnetic waves $\Eb,\Bb$ generated by the  oscillating charge.}  
The IED particle thus defined is geometrically an extensive object, where the term "particle" stands no longer for    a "geometrical point".  Nevertheless, it   maintains the other usual aspects pertinent  of a "particle", such that  it separates one distinct kind of material species from others.
 
The IED process is itself  built on the experimental facts as reflected through a minimal set of established basic, or, first principles laws (see \cite{jxzjied} for a recent review). Based on the first principles solutions for the IED process, it has become feasible to predict a  range of basic properties of particle as demonstrated in [\citeUnif a-m].

It has been one  basic solution obtained earlier [\citeUnif a,c,d] that the total energy $\eng$ of an  IED particle is in direct proportion to the quadratic of the complex total wave function $\psi(x,t)$, i.e. $\eng \propto |\psi(x,t)|^2$, where 
 $|\psi(x,t)|^2$ plays an apparent role of probability  in position space. 
       %field $E(x,t)$, $\eng\propto |E|^2$. With  $E=E_q\psi(x,t)$,  $\psi$ being the dimensionless wave function (the relativistic form of the Schr\"odinger wave $\Psim$) and $E_q$ the amplitude of the field, so  $\eng\propto |\psi(x,t)|^2$. 
             %and thus to $  |\psi(x,t)|^2$ or $|\Psim(x,t)|^2$. 
             %which plays an apparent role of probability  in position space.  
          %or alternatively mass, 
In this paper we shall deduce the $|\psi(x,t)|^2$ as a probability in position space  more rigorously,  and in turn its counterpart in dynamical-variable space,
on the basis of the basic principles of statistics and statistical mechanics, the first-principles solutions for single and many IED particles,  and the   relevant experimental indications. 
%One basic result from the previous  solutions is  that  the IED particle  at the scale of thermal velocity  obeys the Schr\"odinger equation [\citeUnif a,c], and therefore  automatically  the Heisenberg uncertainty principle which was discovered  and wholly proven by W Heisenberg on solid experimental basis\cite{Heisenberg1930}.

\section{Wave mechanics of single IED particle.
Probability density in position space} \label{Sec.Prob}

Consider that  an IED particle  travelling at a velocity $\vel$ along the $x$ axis between two non-absorbing massive walls at $x=0$ and $L$ in the presence of an external  potential field  $\Vcal(x)$. Its constituent point-like charge $q$, as  source, is endowed with a total mechanical energy  (or Hamiltonian) $\eng_q$ at initial time, and will  over time oscillate continuously  at a  characteristic  frequency $\W$, and  also travelling at velocity $\vel$ in the $x$ direction. 
Two  opposite travelling, Doppler-displaced electromagnetic waves of fields $\Eb^j, \Bb^j$, which together with the $q$ constitute the IED particle, are generated by the charge along each direction, where  $j=\dagger$ indicates the component wave travelling in the direction parallel with $\vel$, and $j=\ddagger$ antiparallel with $\vel$. 
              %At the initial time the charge is endowed with a total mechanical energy (or Hamiltonian) $\eng_q$. If the charge had after a finite time ($t_\vphi$)  emitted its entire energy $\eng_q$ with no re-absorption of radiation, then all of the $\eng_q$ is converted to the energy of wave, $\eng=\eng_q$. In the given closed  environment the total energy $\eng_{tot}$ is at any time $t$ carried one portion $a_1$ by the charge, $a_1 \eng_q$, and one portion $a_2$ by the wave, $a_2\eng$; $\eng_{tot}=a_1 \eng_q+ a_2 \eng$.  

In regions excluding the charge,  assuming that, except for 
              %an applied time-independent potential field 
$\Vcal(x)$, no other  charges and currents present, the $\Eb^j, \Bb^j$ fields  are  subject to the Maxwell's equations  
  $$\displaylines{\refstepcounter{equation} \label{eq-Maxwell}
 \hfill \nablab \cdot \Eb^j =0,\quad
 \nablab \cdot \Bb^j=0,   \quad
\nablab \times \Bb^j=0 +\frac{1}{c{'}^2}\pd_t \Eb^j, \quad 
\nablab \times \Eb^j =- \pd_t \Bb^j,  \hfill (\ref{eq-Maxwell})
}$$   
where   $\pd_t \equiv \frac{\pd}{\pd t}$, and
$c{'}^2 = c^2 + \Vcal/m$,   with $c (=\sqrt{1/\ev_0 \mu_0})$   the velocity  of light in a free  vacuum and $m$ the particle's dynamical mass  expressed in (\ref{eq-engEMm}) below. 
Setting $E^j=E_q\vphi^j$, with $E_q$ the amplitude and $\vphi^j$s the (real) dimensionless fields, or wave functions,  their superposition gives the real part of  a  complex total wave function $\psi=E/E_q$,  
 ${\rm Re}[\psi]=\sum_jE^j/E_q =\vphi^\dagsup+\vphi^\ddagsup$.
Further restricting for illustration here the waves to be propagated along the $x$ axis only, we obtain from  the Maxwell's equations a total wave equation $\frac{\pd ^2 \psi }{\pd t^2}=(c^2 + \Vcal(x)/m) \nabla ^2 \psi$. 
The equation, after  routine  transformations under conditions of  small $\Vcal$  and an  insured existence of  stationary-state solutions [\citeUnif c], reduces to 
  $$\displaylines{\refstepcounter{equation} \label{eq-engwav1}
\hfill
 \hat{\eng} \psi =\hat{H} \psi, \quad \hat{\eng}= i\hbar \pd_t, 
\quad \hat{H}= {\hat{p}}^2/m + \Vcal(x), \quad \hat{p}=(\hbar /i)\nabla.
                  % -(\hbar^2/m)\nabla^2 + \Vcal(x).
\hfill (\ref{eq-engwav1})
}$$

If after $N_\vphi$ cycles of continuous oscillations  the charge had emitted its entire initial energy $\eng_q$ without reabsorption, nor with supply of external energy, $\eng_q$ is then converted entirely to the energy of total wave $\psi$  generated by it,  $\eng = \sqrt{\eng^\dagsup \eng^\ddagsup}=\eng_q$.
The total wave $\psi$  extends as a wave train of a mean total length $L_\vphi =\sqrt{L^\dagsup L^\ddagsup}=N_\vphi \lam $.
% and carry  a total wave energy $\eng = \sqrt{\eng^\dagsup \eng^\ddagsup}=\eng_q$.

In the  presence of reflection walls assumed earlier,  the radiation waves $\vphi^\dagsup, \vphi^\ddagsup$, and thus the total wave  $\psi$, will be reflected back from the walls to the charge,  be reabsorbed and re-emitted by it, repeatedly. 
 Assume 
that the particle has reached an equilibrium state in $[0,L]$, with  $L$ being relatively small. Its total energy ($\eng_{tot} $)  will ordinarily be carried in part by  the charge  and in part  by the wave, by the  fractions $a_1$ and $a_2$, with $a_1,a_2 \le 1$, $a_1+a_2=1$, and $\eng_{tot}=a_1 \eng_q+a_2 \eng$. 
The extensivness of the IED particle  is determined by its wave components, the total $\psi $ in question here, and is of our primary concern below.  $\psi$ will be based on in the discussions below.

Our primary concern here  is  "probability".  We thus start by directly writing down  the total electromagnetic energy density, in terms of its total field  $E(x,t)=E_q\psi(x,t)$, in the usual form  $\eng_0(x,t)=\ev_0 |E(x,t)|^2$.   ${\eng}(t)= \int^{\Lw}_0 \eng_0 (x,t) dx=\frac{L_\vphi}{L} \int^L_0  \eng_0 (x,t) dx$ gives the total wave energy at time $t$. It is here that the complex form of $\psi$ finds its clear physical significance: The complex $\psi$ ensures that $\eng$ is a mechanical energy, or, Hamiltonian $\propto |E|^2=({\rm Re} [E])^2+({\rm Im} [E])^2$, which includes both  an inertial component  $\propto ({\rm Re} [E])^2$ carrying the kinetic energy of the field $E$,   and 
 an elastic component $\propto ({\rm Im} [E])^2$ carrying the restoring elastic potential energy of the  vacuum which acquires an induced  shear elasticity in the presence of the  charge $q$ [\citeUnif a,g,h]. 

For stationary state, for which $\eng_0(x,t)$ by definition does not change with time,  
  $$\displaylines{\refstepcounter{equation} \label{eq-enga}
\hfill
\eng_0 (x) 
= \engbar |\psi(x,t)|^2=\engbar |\psi(x)|^2, 
\quad
{\eng}= \int^{\Lw}_0 \eng_0 (x) dx
=\engbar  \int^{L}_0 |\psi(x,t)|^2 d x, 
\hfill
(\ref{eq-enga}) 
}$$
where $\engbar=\Lw \ev_0 E_q^2$ for $\psi $ being normalized in $[0,L]$, 
$$\displaylines{
\refstepcounter{equation} \label{eq-engnorm}
\hfill
\int^{L}_0 |\psi(x,t)|^2 d x=1
\hfill (\ref{eq-engnorm})
}$$
and thus $\eng = \engbar$. The corresponding complex wave function is $\psi(x,t) =\psi(x)e^{-i (\w t +\a_0)}$, where $\w=\g \W$, with $\g=1/\sqrt{1-\vel^2/c^2}$[\citeUnif a,c,d];  
  $\a_0$  is an initial phase which affects the field  configurations of the individual component waves (see further after eqn. \ref{eq-prodens}) but  not the total particle  state  (since  $|e^{i\a_0}|^2=1$) and can be here simply set to $\a_0=0$.  The total linear momentum is $p=\int^L_0 p_0(x) d x$, with $p_0(x) = \frac{(\bar{\eng} -\Vcal(x))|\psi(x)|^2 }{c}$; in the case of $\Vcal(x)=\Vcal_0$ being constant, $p= (\eng -\Vcal_0)/c$.

The $\eng$, in terms of the total wave $\psi$ of a frequency $\w$, is simultaneously subject to the Planck energy equation,
  $$\displaylines{\refstepcounter{equation} \label{eq-Plnk}
\hfill
\eng_{q.n_q}=n_q(b_0/2\pi) \w, \quad n_q=1,2,\ldots, \ b_0=h; \quad 
({\rm for}\ n_q=1): \ \eng_q =\hbar \w, \quad \eng=\eng_q;   
 \hfill (\ref{eq-Plnk})
}$$
(\ref{eq-Plnk}) can be obtained  as a solution to  the energy  equation (internal work), of a form analogous to the Schr\"odinger equation, for the harmonically oscillating   charge $q$  being point like and yet extensive at scale  $\sim 10^{-18}$ m, and of a  dynamical mass ($\Mcal_q$)[\citeUnif c,j]. Furthermore, for the {\it wave train} travelling rectilinearly at a {\it finite} velocity $c$ which resembles a rigid object[\citeUnif a,c,e]  of  a {\it finite} inertial mass ($m$), we may employ  Newton's  law of inertia and obtain $p_0(x)$ in relation to  density of mass, $m_0(x)$, (setting $\Vcal=0$ for defining mass here) and in turn the integration as
$$\displaylines{
\refstepcounter{equation} \label{eq-engEMm}
 \hfill m_0(x)=\frac{p_0(x)}{c}=\frac{\eng_0(x)}{c^2}
=\frac{\engbar}{c^2}  |\psi(x,t)|^2, 
\quad 
m=\frac{\engbar}{c^2} \int^{L}_0 |\psi(x,t)|^2 d x=\frac{\eng}{c^2}=\frac{p}{c}.
\hfill (\ref{eq-engEMm})
}$$
%For the general case of finite $\Vcal$, $M^2c^4 +p_\vel^2c^2 =(\eng-\Vcal_0)^2$, with $p_\vel =m\vel$[\citeUnif c].

If $\psi(x,t)$ were an arbitrary function of $t$, the proportionality  relation $\eng_0 (x) \propto 
|\psi(x,t)|^2$ of (\ref{eq-enga})  would not generally hold, and $\eng$ would be generally dependent on $t$. However under the condition of stationary state  we assumed, (\ref{eq-enga})  holds generally,  as may be readily verified by substituting  $\psi(x,t)=\psi(x)e^{-i(\eng/\hbar)t}$ in the wave equation $\hat{\eng} \psi =\hat{H} \psi$ of  (\ref{eq-engwav1}), and multiplying it by $\psi^*$  from the left. Its left side gives 
            %the $\eng_0(x) $ of 
(\ref{eq-enga}a), or rewritten as 
 $$\displaylines{\refstepcounter{equation} \label{eq-densop}
\hfill
\eng_0(x)
=  \psi^* \hat{\eng} \psi=\engbar |\psi(x,t)|^2 \quad (a), \quad \eng = \int^L_0 \psi^* \hat{\eng} \psi dx =
\int^L_0 \psi^* \hat{H} \psi  d x \quad (b)
\hfill (\ref{eq-densop})
}$$  
where  (\ref{eq-densop}b) is given by integrating  (\ref{eq-densop}a) on both sides.

For the specific case of  $\Vcal(x)=0$ and $\vel=0$, (\ref{eq-engwav1}) has the solution $\psi=\frac{1}{\sqrt{L}}e^{i(K x-\W t )}$, where $K=\lim_{\vel^2/c^2\rightarrow 0}k(=\w/c)=Mc/\hbar$, $\W =\lim_{\vel^2/c^2\rightarrow 0} \w= Mc^2/\hbar$ and $M=\lim_{\vel^2/c^2 \rightarrow 0} m$[\citeUnif a,c-e]; with this $\psi$ in  (\ref{eq-engwav1}) we obtain $\hbar \W\psi = \frac{\hbar^2K^2}{M}\psi$. Subtracting the resulting equation from (\ref{eq-engwav1}) of a generally finite $\Vcal$ and $\vel$, taking limit at $\vel^2/c^2\rightarrow 0$ and at  scale of the de Broglie wavevector $k_d=(\frac{\vel}{c})k$ ($<<k$), $\Psim=\lim_{\vel^2/c^2\rightarrow 0, k_d/K \ll 1 } \psi= \Psim(x)e^{-i \Wvel t}$, with $\Wvel=\frac{1}{2}\frac{\vel^2}{c^2}\W$,
we obtain a Schr\"odinger form of wave equation governing the particle's kinetic motion [\citeUnif b,c], $\hat{\Eng}_\vel \Psim = \hat{H}_\vel\Psim $,
where $\hat{\Eng}_\vel=\hat{\eng}-\hbar \W $, $\hat{H}_\vel= \hat{H}-\frac{\hbar^2K^2}{M}=\frac{ \hat{P}_\vel^2}{2M} +\Vcal$, and $\hat{P}_\vel=\frac{\hbar }{i} \nabla$. 
The corresponding energy density and integration follow to be  
  $$\displaylines{\refstepcounter{equation} \label{eq-engv0}
\hfill
\Eng_{\vel 0}(x) =\Psim^*\hat{\Eng}_\vel \Psim= \bar{\Eng}_{\vel} |\Psim(x,t)|^2, 
\quad  
  \Eng_\vel =\int^L_0 \Psim^* \hat{\Eng}_\vel \Psim dx =\int^L_0 \Psim^* \hat{H}_\vel\Psim dx, 
\hfill (\ref{eq-engv0})
}$$ 
where $\bar{\Eng}_\vel =\Eng_\vel$ for  $\int^L_0 |\Psim|^2dx=1$; for the  case of $\Vcal(x)=\Vcal_0=$constant,  
$\Eng_\vel=\frac{1}{2}M \vel^2+\Vcal_0$.
%=\hbar \Wvel$. 
We shall  proceed  the remainder of the discussions in terms of the  total wave $\psi$ and energy $\eng$, etc., i.e., the  "relativistic" forms, and in one dimension. The conclusions can all be  carried in a straightforward way over for $\Psim$, $\Eng_\vel$, etc. by taking the limit at  $\frac{\vel^2}{c^2}\rightarrow 0$ and scale $k_d$, and to three dimensions[\citeUnif c].

 Unambiguously, the total energy or  mass  of the IED particle is as any  real particle  an absolute measure of the existence of the particle. The IED particle of a spatially extensive energy and mass as of (\ref{eq-enga}), or (\ref{eq-densop}), and (\ref{eq-engEMm}) therefore is extensive, here apparently  across [$0,L$] (though practically across $L_\vphi$). And its  presence at position $x$ in [$0,L$]  is measured by the energy density  $\eng_0(x)$, or alternatively  mass density  $ m_0(x)$, which for stationary state contains each the common factor $|\psi|^2$, denoting by $\rho_{_l}(x)$  as 
$$\displaylines{
\refstepcounter{equation} \label{eq-prodens}
\hfill
\rho_{_l}(x)=|\psi(x,t)|^2, \quad 
 \hat{\Kcal} \rho_{_l}  = \psi^*(x,t) \hat{\Kcal} \psi(x,t) = |\psi(x,t)|^2 \overline{\Kcal}.
\hfill (\ref{eq-prodens})
}$$ 
where in  (\ref{eq-prodens}b) $\rho_{_l}(x)$ is written  in the more general form of being acted on by a dynamical variable operator $\hat{K}$.

The foregoing descriptions relevant to $\rho_{_l}(x)$  may be summarized as:  (i) $\rho_{_l}(x)$ is a distributed function in $[0,L]$ describing  electromagnetic fields which are implicitly random because the initial phase  $\a_0$ of $\psi$ (see after eqn. \ref{eq-engnorm}) is  in general the result of  random  interactions (Sec. \ref{Sec-ProbMomnSpc}). (ii) $\rho_{_l}(x)$ is a measure of the  fraction of the total energy or mass of  the particle presenting at  $x$. And $\eng$ and $m$ are the  $\rho_{_l}(x)$-weighted integrations across $[0,L]$, and therefore the expectations of total energy and mass. (iii) $0\le \rho_{_l}(x)\le 1$ according to (\ref{eq-enga}) and (\ref{eq-engnorm}); in particular, $\rho_{_l} (x)=0$ if no portion of the particle is found at  $x$, and $\rho_{_l}(x)=1$ if the entire particle is  found at $x$, hence  a corpuscle. 

Points (i)-(iii) furnish
$\rho_{_l}(x)=|\psi(x,t)|^2$, or $\rho_{_l}(x)\dot{=}|\Psim(x,t)|^2$ at scale $k_d$,
with all of the essential qualifications for $\rho_{_l}(x)$ 
%or $\rho_{_l}(x)\dot{=}|\Psim(x,t)|^2$ at scale $k_d$, 
to be defined as  the probability, or probability density here, for finding a corresponding portion $\rho_{_l}(x)$ of  the IED particle at position $x$ in a given stationary state. This conclusion is in complete accord with the (mathematical) probability assumption in the formal theory of quantum mechanics.

\section{Time evolution of single particle state. 
Probability density in dynamical variable space
}\label{Sec-ProbMomnSpc}

Let now the $N=1$ IED particle in the one-dimensional box of volume $V=L$ be  subject to interactions with  its environment,  the wall particles and/or environmental radiation fields, maintained at a constant  temperature $T$. The particle is here described in the quantum-mechanical terms by the  wave equation (\ref{eq-engwav1}) for  all regimes, and the description will reduce to  the classical mechanics limit when the particle's energy levels  $\eng_\nu$s are on the scale $h$ densely nested with one another. 
                %or $\Delta \eng_\nu>>\hbar /\tau$.
Assume that the interactions are weak, infrequent,  and elapse over a brief time $\delta t_\nu$ only on each occurrence. 
So the IED particle will maintain a given stationary  state $\nu$ (specified by e.g. a total energy, linear momentum, and spin, etc.) for a finite time interval $t_{\nu}, >>\delta t_\nu$, but,  subjected to  interactions  with  environment,   will explore all possible states on a time scale $t>>t_\nu$.
% after  sufficiently  long time. 
               %compared to the relaxation time of the system. 
                %********************
                % cf p58, D Chandler
                %********************

This may for example describe the realistic case of an electron moving in a fixed orbit about  the nucleus of  an isolated atom until an  emission/absorption of  radiation, or a conduction electron  moving at a  constant velocity in a small metal specimen until next collision, or a spin-half atom  traversing a magnetic field with a fixed spin orientation  as in the Stern-Gerlach experiment (Sec. \ref{Sec-ProbMomnSpc}) until externally perturbed. Under the condition of $t_{\nu} >>\delta t_\nu$, contributions from transient states may be neglected, and  
the thermodynamic properties are essentially determined by the stationary states.

Instead of making observations over time, we now  "compact" a large   $\Nstat$ number of the states evolved over  time ($t'$) into $\Nstat$ replicas which each describe the  one particle system here under condition of constant $T,V,N(=1)$, and are uniformly distributed over a total $\Nstat$ number of possible quantum states of the particle.
          %and simultaneously evolve in time ($t$).
 We thus obtain a canonical ensemble. Each $\nu$th (eigen) state is associated with an (eigen) wave function $\psi_\nstat(x,t)=\psi_\nu(x)e^{-i(\eng_\nu/\hbar)t}$, energy $\eng_\nu$, etc., given as the solution of (\ref{eq-engwav1}), with $\nu=1,2, \ldots, \Nstat$. Each $\nu$th state occupies a volume element in the phase space spanned here by $x,p_x$; the trajectory of state in this space corresponds to the time evolution of state of the actual particle. With (\ref{eq-engwav1}) being a linear equation,  the linear sum of $\psi_\nstat(x)$s, 
$$\displaylines{
\refstepcounter{equation} \label{eq-Psimt}
\hfill 
\psi_{\ens}(x,t) = \sum_{\nst=1}^{\Nstat}  c_\nst(t)        \psi_\nst(x) 
\hfill
(\ref{eq-Psimt})
}$$
must  again be a solution of  (\ref{eq-engwav1}),  where  $c_\nu(t)$ is a  "weight coefficient" or "amplitude" of state $\nu$ and is in general complex.

That $\psi_{\ens}(x,t)$ is a linear sum of $\psi_\nu$s   implies only the linearity of the   wave equation.  $\psi_{\ens}(x,t)$  does not describe the total displacement at $x$ due to the superposition of $\psi_\nu$s, since the $\psi_\nu$s   
occur in reality at different times and never meet. 
               %**********keep :******************************************
               %       \footnote{(\ref{eq-Psimt}) has an analogous function form to the standard QM aimed to describe  a wave packet  (e.g. \cite {Merzbacher:1970}, p47), a picture that is however compelled to be abandoned here since it for example can not lead to the experimentally observed self interference of practically any interference phenomena of particles [\citeUnif k].)} 
                 %*********************************************************

Multiplying $\psi_{\nst'}^*(x) $ on  both sides of (\ref{eq-Psimt}) from the left   and integrating over  $(0,L)$, we obtain, for the $\psi_\nst$s being orthonormal ($\int^L_0 \psi_{\nu'}\psi_\nu dx =\delta_{\nu',\nu}$, $\delta_{\nu',\nu}$ being the Kronecker delta), 
$$\displaylines{
 \refstepcounter{equation} \label{eq-cnup}
\hfill
\int^L_0 \psi_{\nst}^*(x) \psi_\ens(x,t) d x = \int^L_0 \psi_{\nst}^*(x)  \sum_{\nu'} c_{\nu'}(t) \psi_{\nu'}(x)  dx
= c_{\nst} (t). 
\hfill  (\ref{eq-cnup})
}$$
For the canonical ensemble of a dimension $\Ncal$ assumed large, its average energy, proportional to the integral below, is evidently a constant. And the integral itself must aslo be a constant, and is  normalised to one as 
$$\displaylines{
\refstepcounter{equation} \label{eq-psiennrm}
\hfill
\int^L_0 \psi_\ens^*(x,t) \psi_\ens(x,t) d x 
 =1, \quad {\rm or} \  \sum_{\nst}^{\Nst}  c_\nst^*c_\nst \int^L_0  |\psi_{\nst}(x)|^2 d x
=  \sum_{\nst}^{\Nst}  |c_\nst(t)|^2 =1,
\hfill (\ref{eq-psiennrm})
}$$
where $|c_\nst(t)|^2 =c_\nst^*(t) c_\nst(t)$. With $\psi_\nst$ for $\psi$  in (\ref{eq-engwav1}), multiplying on both sides of the  $\nu$th equation  by $c_\nst$ from the left  for  $\nu=1, \ldots,\Ncal$, summing the equations over all $\nu$, we obtain a corresponding  wave equation,  
$$\displaylines{
\refstepcounter{equation} \label{eq-waveqens}
\hfill
\hat{\eng} \psi_\ens =\hat{H} \psi_\ens, 
\quad 
\hat{\eng} =i \hbar \pd_t, 
\quad 
\hat{H} = -(\hbar^2/m) \nabla ^2  + \Vcal(x).  
\hfill (\ref{eq-waveqens})
}$$

With $\psi_\nu$ for $\psi$,  making similar algebraic operations as  in leading to  $\eng$ of (\ref{eq-densop}), we obtain the  energy of $\nst$th state, $\int^L_0 \psi_\nst^*(x,t)i\hbar \frac{\pd \psi_\nst(x,t)}{\pd t} dx =\eng_\nst $. 
From the significance of $c_\nst$ as the amplitude of state $\nu$ in $\psi_\ens$ defined in (\ref{eq-Psimt}), the energy expectation  over the states of the ensemble follows to be  $\langle\eng_\nst\rangle =\int^L_0  \psi^*_\ens(x,t) i \hbar  \frac{\pd \psi_\ens(x,t)}{\pd t} dx $,  and 
$$\displaylines{
\refstepcounter{equation} \label{eq-engens}
\hfill 
\langle\eng_\nst\rangle
=\int^L_0 \sum_{\nst'=1}^{\Nstat}c_{\nst'}^*(t) \psi_{\nst'}^* (x) 
 \sum_{\nst=1}^{\Nstat} \eng_\nst c_\nu(t) \psi_\nst(x) dx
=\sum_{\nst=1}^{\infty}  \eng_\nst \rho_\nst, 
\quad \rho_\nst= |c_\nst(t)|^2,
\hfill (\ref{eq-engens})
}$$
where for the final expression for $\langle\eng_\nst\rangle$, $ \int^L_0 |\psi_\nst|^2 dx=1 $ is used. The ensemble average $\langle\eng_\nst\rangle $ of  (\ref{eq-engens}) is equivalent to the time average provided that the  IED particle here is {\it ergodic}, an assumption usually made  for  real particles. 
  %******
  %cf p58 D Chandler       ********
  %*******
By a similar reasoning as for $\rho_{_l}(x)$ earlier (as $\rho_{{_l}_\nu}(x)=|\psi_\nu(x,t)|^2$ here), the  $\rho_\nst=|c_\nst(t)|^2$ given in (\ref{eq-engens})  has all of the qualifications to be defined as  the probability of finding the particle  at  state $\nst$. The function $\rho_{\nst} $ is related to $\eng_\nu$ through a Boltzmann factor, $ \propto e^{-\eng_\nst/k_B T}$ which is a proper problem of statistical mechanics.

 Given an arbitrary dynamical variable $K_\nu(x)$ of an operator $\hat{\Kcal}_\nu(x)$ (which may also be an  independent variable like the $x$, or  the $\sigma$ later)  of the single particle for $\nu$th state at position $x$, from the preceding discussions its expectation 
in both the position and dynamical-variable  spaces follow to be  
$$\displaylines{
 \refstepcounter{equation} \label{eq-Knav}
\hfill
\langle \Kcal\rangle
= \hspace{-0.2cm}\int_{0}^{L} 
\sum_{\nu=0}^{\infty} c_\nu^*\psi_\nu^* \hat{\Kcal}_\nu(x)
    \sum_{\nu'=0}^{\infty} c_{\nu'} \psi_{\nu'} d x  
= \sum_{\nu=0}^{\infty} |c_\nu|^2  \overline{\Kcal}_\nu, 
\quad 
\overline{\Kcal}_\nu=\hspace{-0.2cm}\int^{L}_{0} 
\psi^*(x) \hat{\Kcal_\nu} \psi_\nu(x)dx. 
\hfill (\ref{eq-Knav})
}$$

The  $\psi_\nu$s have been introduced in the above as describing events taking place  in reality in a random fashion one after another in  time, only {\it one at a time} regardless of being probed by an observer or not. This feature is in direct accord with the eigen-state solutions of  Schr\"odinger equation for a corresponding system and, as  can be expounded case by case, with overall experiments.  

A most clear-cut experimental indication  is provided by the Stern-Gerlach experiment ({\it Z Phys} {\bf{8}}, 110, 1921). 
             %******************
            % W Stern and O Gerlach, Z Phys \bf{8}, 110 (1921)  
             %******************
Here, a silver atom, of a net spin $\frac{1}{2}$ due to its outermost-shell electron
and being ordinarily random oriented, is let travel at a constant velocity ($\vel$) in the $x$ direction  across a region $d$ applied with  an inhomogeneous  transverse ($z$-direction) magnetic field $B_z$ of a finite gradient $\frac{\pd B_z}{\pd z} (> 0)$. In the $B_z$ field  its spin magnetic moment $\mu_z=-\frac{\sigma}{2}g_s \mu_{\Bsub}$ is quantized in orientation (Zeeman effect), $\sigma =+1$ or $-1$. If the atom deterministically maintains a fixed spin  orientation, say  $\sigma=$  $+1$, provided also that no other perturbation presents,  the eigen  state function being thus $\psi=\psi_{+}$, then across the entire $d$ region the atom will constantly be acted by a force $F_z = \frac{1}{2} g_s \mu_{\Bsub} \frac{\pd B_z}{\pd z}$  in the fixed $+z$ direction for a duration $t=d/\vel$, and at the exit of $d$ be displaced  by a finite distance  $+ z_1 =\frac{1}{2}\frac{|F_z|}{m} t^2$ in the $+z$ direction (in the experiment the atom also traverses an additional free flight distance $D$ which is not of direct concern here). Or if  $\sigma=-1$, the  state function being thus $\psi=\psi_{-}$, then  $F_z$ and $ z_1$ change each to  the $-z$ direction. The above is  as observed in the experiment. 

On the contrary, if during each one trip the atom  were in a mixture of both spin up and down states, so $\psi=\frac{1}{2}\psi_{+}+\frac{1}{2}\psi_{\mbox{-}}$, then across $d$ the atom would be acted by  forces of alternating directions. The so  produced  net displacement at the exit must on average be zero, $ z_1=0$, which is in contradiction with the experimental observation.

\section{The state vector space}
\label{Sec-LinVect}

What is mainly new in the foregoing is the concretisation of the physics underlining  the probabilities in both the position and dynamical variable spaces instrumented  by the IED model and solutions. The involved mathematical structure and operations are  of no essential difference from the usual quantum mechanics,   since the governing equations  at scale $k_d$ are the  common  Schr\"odinger equation. This feature holds  through the remainder of discussions. 

As in the unified mathematical formulation of quantum mechanics, the description in the (position) coordinate space can be transformed to one in an  $\Ncal$-dimensional linear state vector space (the Hilbert space for  $\Ncal$ being infinite),   which for the single IED particle  here is spanned by the $\psi_\nu$s as coordinate axes, or basic vectors,  of a total number $\Ncal$,  assuming the $\psi_\nu$s orthonormal in $(0,L)$.
 For many (IED) particles, we have the Fock space.
The  linear combination of the $\Ncal$ basic vectors $\psi_\nu$s,  $\psi_\ens=\sum_\nu c_\nu \psi_\nu$ of  (\ref{eq-Psimt}), represents in this space a "(general)  state" vector  which describes based on  Sec. \ref {Sec-ProbMomnSpc}  the state of the (canonical) ensemble; $c_\nu$ is a component of the vector $\psi_\ens$. And $|c_\nu|^2 $ is the probability of "finding" the physical system, the single particle in $(0,L)$ here, in $\nu$th (stationary) state at a time point $t'_\nu$ on the time evolution axis $t'$. 

With the $\psi_\nu$s being discrete,  the linear mathematical  operations in the space spanned by the $\psi_\nu$s may be naturally carried out  in terms of matrix analysis.  
If for example  $\Kcal_{a i},\Kcal_{b j}$ ($i,j=1,2, \ldots, \Ncal$) are given as two dynamical variables,   their two general states are  given by $\psi_{\ens.a}(x,t)=\sum_i a_i(t) \psi_{i} (x) $, $\psi_{\ens.b}(x,t)=\sum_j b_j(t)  \psi_{j} (x)$. 
 Their scalar product for example is  defined by   
$$\displaylines{\refstepcounter{equation} \label{eq-abm}
\hfill 
(\psi_a,\psi_b)
=\int^L_0 \sum_\nu  a_\nu^* \psi^*_\nu  \sum_{\nu'}  b_{\nu'} \psi_{\nu'} dx =\sum_\nu a_\nu^* b_\nu 
= (a_1^*,a_2^*, \ldots, a_\Ncal^*)\left(\begin{array}{c}
b_1   \cr
b_2 \cr
\vdots\cr
b_\Ncal 
\end{array}\right)= \langle a|b\rangle,           \hfill (\ref{eq-abm})
}$$
where $\int^L_0 \psi^*_\nu\psi_{\nu'} dx=\delta_{\nu \nu'}$ is used; $(\psi_a = \langle a|$ corresponds in the  final expression to a  $ \Ncal \times 1$ matrix with its  $\Ncal$   elements  being imaginary and arranged in  one raw,  and $\psi_b)=|b\rangle$ a  $1 \times \Ncal$ matrix with its $\Ncal$   elements being real and in one column. If $a=b=c$, (\ref{eq-abm})  returns the  result of  (\ref{eq-psiennrm}).
The operator $\hat{H}$ or $\hat{\Kcal}$ in general is an $\Ncal \times \Ncal $ matrix.

\section{$N$ weakly interacting  particles in a mean interaction field} 
\label{Sec-ProbMany}

Let now  $N$  weakly interacting    IED  particles of the  dynamical masses $m_a, m_b, \ldots$, wave functions $\psi_a(x_1,t), \psi_b(x_2,t), \ldots$, and attached with the coordinates $x_1, x_2, \ldots$, be enclosed in the box of size $L$ in an effective mean potential field $\Vcal(x_1,x_2, \ldots,x_\Nsub), =\frac{1}{N(N-1)}\sum_{i=1}^{N} \sum_{j=1, (j\ne i)}^{N-1} \frac{1}{2}\Vcal_{ij}$, with $\Vcal_{ij}$ a pairwise interaction potential. Each $\psi_\a(x_i,t)$ is governed by a wave equation given after (\ref{eq-engwav1}),
$$\displaylines{
\refstepcounter{equation} \label{eq-waveqa}
\hfill 
\hat{\eng}_\a  \psi_\a(x_i,t) =\hat{H}_{ \a}\psi_\a(x_i,t), 
                 \quad         \hat{\eng}_\a  =  i \hbar \pd_t,
            \quad \hat{H}_{\a}= -(\hbar^2/m_\a) \nabla_i^2 +\Vcal(x_1,\ldots, x_\Nsub), 
\hfill (\ref{eq-waveqa})
}$$
where $\a=a, b, \ldots, \a_\Nsub,  \ i=1,2, \ldots, N$. The $N$  $\psi_\a $-waves  are each distributed across the entire $L$, thereby overlapping with one another. The total field at  position  $x_i$ is $\psi_\Sigma(x_i,t)=\sum_\a \psi_\a(x_i,t) $, which generally contains  a finite portion of each of the $N$ waves, or, particles.
From a measurement of $\psi_\Sigma(x_i,t)$ at $x_i$,  we are thus not  able to distinguish which particle this is; that is, the  $N$ particles are indistinguishable, and  in E. Schr\"odinger's terminology  ({\it Proc Cam Phil Soc} {\bf 31}, 555,
1935), they are {\it entangled} with one another.
                   %***********************************keep************************************
                  %\cite{schordinger1935}: discussion of probability relations between separated systems, Proc Camb Philos Soc {\bf 31},  555, 1935
                   %***********************************keep************************************

All of the expressions up to  (\ref{eq-abm}) can accordingly be written down for each $\a$th particle. Of specific interest here is   the probability density of finding particle $\a$ at $x_i$, $\rho_{{_l}_\a}(x_i) =|\psi_\a(x_i,t)|^2$  given after (\ref{eq-prodens}), with $\a=a,  \ldots, \a_N,  \ i=1, \ldots, N $.   The notion of $|\psi_\a(x_i,t)|^2$ as a probability  may be advanced  for the many-particle system  through excising it in a few representative cases below, as we shall see, each returning the expected result of the formal quantum mechanics and (or) experiment. 
                   %***********************************keep************************************
                   %(see e.g. p1246, survey of applicable mathematics, by K rektorys 1969),
                   %***********************************keep************************************

(i)  The  probability  of the simultaneous occurrence of  $N$ independent events that particle  $a$ is at $x_1$, $b$ is at $x_2$, $\ldots$  and $\a_\Nsub$ is at $x_{\Nsub}$ with the  probabilities $\rho_{{_l}_a}(x_1), \rho_{{_l}_b}(x_2), \ldots$ and $\rho_{{_l}_\Nsub}(x_{\a_\Nsub})$, with the particles being in the mean field effectively non-interacting, is according to  the usual multiplication rule for probabilities given by 
$$ \prod_{       \{\a,i\}=\{a,1\}  }^{\{\a_\Nsub, N\}}  \rho_{{_l}_\a}(x_i) =|\psi_a(x_1) \psi_b(x_2) \ldots \psi_{\a_\Nsub}(x_\Nsub)|^2.$$
 Because of  indistinguishablity, making  simultaneous pair-wise coordinate  permutations say from $x_1 $ to $x_{2}$ for particle $a$ and $x_{2}$ to $x_1$ for $b$, leaves the product $|\psi_a(x_2) \psi_b(x_1)|$
 unchanged except for a  change in sign if the permutations are antisymmetric; 
there are a total $N!$ number of possible permutations.  To ensure an always  positive probability we should introduce as in usual practice a "Sign" into the product of $N$ wave functions as  (Sign)$\psi_a(x_1) \psi_b(x_2) \ldots \psi_{\a_\Nsub}(x_\Nsub)$. 
Further dividing $\sqrt{N!}$ for normalisation, the complete equation for the probability is therefore   
$$\displaylines{\refstepcounter{equation} \label{eq-proba}
\hfill 
\rho_{{_l}_\Nsub} (x_1, x_2,\ldots,x_{\Nsub})
=\sum_{i,i'}^{N!} | \psi_\Nsub(x_1,x_2, \ldots, x_\Nsub; t) |^2
= \sum_{i,i'}^{N!} \prod_{       \{\a,i\}=\{a,1\}  }^{\{\a_\Nsub, N\}}  \rho_{{_l}_\a}(x_i),
\hfill (\ref{eq-proba})
\cr
{\rm where} \hfill
\cr
 \refstepcounter{equation} \label{eq-probpsi}
\hfill 
 \psi_\Nsub(x_1,x_2, \ldots, x_\Nsub; t) 
=({\rm Sign})\frac{1}{\sqrt{N!}}\psi_a(x_1,t)\psi_b(x_2,t) \ldots \psi_{\a_\Nsub}(x_\Nsub,t), 
\hfill (\ref{eq-probpsi})
}$$ 
with  Sign$=+1$  for $ \psi_\Nsub$  symmetric and $-1$  antisymmetric in respect to simultaneous  permutations between a pair of  coordinates $x_i$ and $x_{i'}$.  

One common application of the function form $\psi_\Nsub$, a complex  $N$-particle wave function, is the description of  the simultaneous states of many electrons in an  atom, or in the atoms of a  condensed matter. Multiplying (\ref{eq-waveqa}) by $\frac{1}{\psi_a(x_i,t)}$  from the left, summing the equations (\ref{eq-waveqa}) over all $\a$s, reorganising, we obtain  a wave equation  for $\psi_\Nsub$ as 
$$\displaylines{\refstepcounter{equation} \label{eq-wavpsiNs}
\hfill
\hat{\eng}_\Nsub \psi_\Nsub=\hat{H}_{ \Nsub}\psi_\Nsub, 
            \quad 
\hat{\eng}_\Nsub =i\hbar \pd_t, \quad 
\hat{H}_{\Nsub}= -\sum_{(i,\a)=(1,)}^{(N,\a_\Nsub)}\frac{\hbar^2}{m_\a} \nabla_i^2 +N \Vcal. \hfill (\ref{eq-wavpsiNs})
}$$

(ii) A beam of  ($N$) incoherent identical particles, each being a travelling wave  $\psi(x_i,t)$,  is incident from left onto an opaque screen with two slits $A,B$, and emerges from these  as $\psi_\Asub(r_1,t), \psi_\Bsub(r_2,t)$, which are then recorded at a point $P$ on 
a photographic plate on the right. 
We register  the passings of the $\Ncal=2$ slits $A$ and $ B$ by the incident beam that have subsequently arrived at $P$  
as two events, which are here mutually exclusive and  have  the probabilities  $\rho_{{_l}_\Asub}(P)=|\psi_\Asub (P,t)|^2$ and $\rho_{{_l}_\Bsub}(P)=|\psi_\Bsub(P,t)|^2$. 
 The probability of  the occurrence of at least one of the $\Ncal=2$ 
two mutually exclusive events 
    %that  at least one of the   $\Ncal=2$ slits $A$ and $ B$ is  passed by the  particle beam having later arrived at $P$, with the probabilities  $\rho_{{_l}_\Asub}(P)=|\psi_\Asub (P,t)|^2$ and $\rho_{{_l}_\Bsub}(P)=|\psi_\Bsub(P,t)|^2$, 
is according to the usual addition rule for probabilities given by 
$$\displaylines{\refstepcounter{equation} \label{eq-probsum}
\hfill
\rho_{{_l}_{\Asub\Bsub}}(P) =|\psi_\Asub (P,t)|^2+|\psi_\Bsub (P,t)|^2.
 \hfill (\ref{eq-probsum})
}$$

(iii) If the incident  particle  beam of  (ii) consists of ($N$) coherent waves, then the $\Ncal=2$ events are no longer mutually exclusive, but are in a situation as stated by  the usual conditional probability ($\rho_{\Asub\Bsub}
=\frac{1}{2}(\rho_\Bsub\rho_{\Asub/\Bsub}+\rho_\Asub\rho_{\Bsub/\Asub}$)); the expressions for $\rho_{\Asub/\Bsub}$ etc. are subject to how the events specifically depend on one another. Concretely, for the present case we can readily write down the total wave displacement at $P$ as $\psi_{\Asub \Bsub}(P,t)= \psi_{\Asub} (P,t) + \psi_{\Bsub} (P,t)$. But $|\psi_{\Asub\Bsub}(P,t)|^2$ is according to Sec. \ref{Sec.Prob} just the probability $\rho_{{_l}_{\Asub\Bsub}}(P)$, so 
$$\displaylines{\refstepcounter{equation} \label{eq-probsumc}
\hfill
\rho_{{_l}_{\Asub\Bsub}}(P)=|\psi_{\Asub\Bsub}(P,t)|^2=| \psi_{\Asub} (P,t) + \psi_{\Bsub}(P,t)|^2.
\hfill (\ref{eq-probsumc})
}$$ 
This will predict  the interference fringes observed in double slit experiments. For $N$ identical particles as in  (ii)-(iii) above,  $m_\a=m$ for all $\a$s; directly summing the (\ref{eq-waveqa})s over all $\a$s gives a total wave equation for $\psi_\Sigma$ as 
$
i \hbar \frac{\pd  \psi_\Sigma(x_i, t) }{\pd t}=\hat{H}_i \psi_\Sigma(x_1,t)$, 
$\hat{H_i}= - \frac{\hbar^2}{m} \nabla_i^2 + \Vcal. $

%\subsection*{Acknowledgements}
%\ack

The author expresses thanks to 
           %Professor H.-D. Doebner for his invaluable discussions over this work at the 28th Int Colloq Group Theo Meth Phys (Group 28, Univ Northumbria),  extending our previous discussions during annual meetings  since 2005 and a visit to him in Munich in 2008, which have  benefited clearer presentations of the IED model and solutions,  to 
                 %
%the conference organisers and  chairmen Professors  M Angelova and W Zakrzewski  for arranging the presentation of this work at the 28th Int Colloq Group Theo Meth Phys (Group 28, Univ Northumbria), 
  Scientist P.-I. Johansson for his continued moral and private funding support for the research, and to a community of national and international distinguished  physicists for giving their invaluable moral support for the research. 
             %to  BJ, MHJ, NA for invaluable  moral support of the research, 

% \appendix


\section*{References}
\begin{thebibliography}{5}

\bibitem{Merzbacher:1970}
 Dirac P A M  1958 {\it The Principles of Quantum Mechanics}, 4th ed, (Oxford: Clarendon Press); Weyl H 1931 {\it The theory of groups and quantum mechanics}, English  trans. H P Roberton  (Methuen Comp Ltd);  Merzbacher E 1970  {Quantum Mechanics}, 2nd ed, (John Wiley \& Sons, Inc.). 
                %p36 : "we have regarded the wave function as 'a measure of the probability '  of fing the partcile at time t at the position r"p295 

            %\bibitem{Bohr1920} Bohr N 1920 \"Uber die serienspektra der element {\it Zeit.f. Physik {\bf 2}  423-478,   Bohr N  1976 The correspondence principle (English trans.) {\it Niels Bohr Collected Works}, vol. 3,  eds. Rosenfeld, L.; Nielsen, J. Rud (Amsterdam: North-Holland) pp. 241-282. 


\bibitem{Born1926} Born M 1926 Zur Quantenmechanik der stoßvorg\ss nge, {\it Zeit fur Physik} {\bf 37}, 863-867;
         %. doi:10.1007/BF01397477.              
          %http://www.springerlink.com/content/h06w8465t710u328/. Retrieved 2008-12-16.
         %-Quantum Mechanics of Collision M. Born Z
        %Max Born,
1954 The statistical interpretation of quantum mechanics {\it Nobel Lecture}.
       %, (11 Dec., 11, 1954).


%\bibitem{Planck1900} M Planck, "On the theory of the energy distribution law of the normal spectrum", read at a meeting of the German Physical Society, Dec 14, 1900
%\bibitem{deBroglie1923} De Broglie, 1923
%\bibitem{Dirac1928A}


\bibitem{Heisenberg1930} Heisenberg W 1927 "\"Uber den anschaulichen inhalt der quantentheoretischen kinematik und mechanik, {\it Z. Phys.} {\bf 43} 172;
                          %%?Dokumente der Naturwissenschaft 4 (1963) 9.
              %("The Physical Content of Quantum Kinematics and Mechanics", English trans  in: {\it Quantum Theory of Measurement}, ed. by J. A. Wheeler, W. H. Zurek, Princeton University Press, N.J. (1983) 62.);  
1930 {\it The Physical Principles of the Quantum Theory}, C. Eckart and F. C. Hoyt German to English transl, (Univ. Chicago Press).
              % xx, 1927 ?


\bibitem{jxzjied} Zheng-Johansson J X 2010 Internally electrodynamic particle model: its experimental basis and its predictions {\it Phys. Atom. Nucl.} {\bf 73}  571-581  ({\it Preprint} arxiv:0812.3951).
%\bibitem{jxzj-qho}Zheng-Johansson JX, "harmonic oscillator" 

\bibitem{Unif1} \label{Ref1} Zheng-Johansson J X and Johansson P-I  
(a) 2006 {\it Unification of Classical, Quantum and Relativistic Mechanics and of the Four Forces} 
(New York: Nova Science);  {\it Preprint} arxiv:physics/0412168; 
                    %1st  printing, 2006aii;
(b) 2010 {\it Inference of Basic Laws of Classical, Quantum and Relativistic Mechanics from First-Principles Classical-Mechanics Solutions}   (New York: Nova Sci.); 
          %ISBN: 1-59454-260-0; 
          %ISBN 1-59454-261-9.
          %see physics/0412168(b) for a primordial version;             
          %physics/0501037 (c); 
           %"Inference of Schr\"odinger Equation from
          % Classical-Mechanics  Solution", 
          %\bibitem{xzj-vacdiel} 
(c) 2006 
 Inference of Schr\"odinger equation from classical mechanics solution 
 {\it Quantum Theory and Symmetries IV} {vol 2} {\it Suppl. Bulg. J. Phys.} {\bf 33}, ed  Dobrev VK
(Sofia: Heron Press)  pp 763-770  ({\it Preprint}  arxiv:physics/0411134v5);
                    %"Schr\"odinger Equation for Electrodynamic Model Particle," (internal); 
(d) 2006 Developing de Broglie wave 
{\it Prog. Phys.} {\bf 4}  32-35 ({\it Preprint} arxiv:physics/0608265);  
(e) 2006 Mass and mass--energy equation from classical-mechanics solution
{\it Phys. Ess.} {\bf 19} 544 
({\it Preprint} arxiv:physics/0501037);
(f) (Zheng-Johansson J X) 2006 
Spectral emission of moving atom 
{\it Prog.  Phys.} {\bf 3} 78-81({\it Preprint}  arxiv:physics/060616);
(g) (Zheng-Johansson J X) Vacuum structure and potential
{\it Preprint} arxiv:physics/0704.0131;
(h) (Zheng-Johansson J X) 
Dielectric theory of the vacuum   {\it Preprint} arxiv:physics/0612096; 
 (i)   (Zheng-Johansson J X, Johansson P-I , et al )          2006   Depolarisation radiation force  in a dielectric 
               medium. its analogy with gravity 
{\it Quantum Theory and Symmetries IV} vol 2,  {\it Suppl. Bulg. J. Phys.} {\bf 33}
              ed Dobrev V K 
(Sofia: Heron Press),  pp. 771-779; 
  (Zheng-Johansson J X, Johansson P-I)  arxiv:physics/0411245v4; 
             %"Depolarisation Radiation Force in a 
             %Dielectric medium. Its Analogy with Gravity", 
             %pp. yy-yy; 
             %\bibitem{jxzj-mass}
               %(n): (J.X. Zheng-Johansson), "Inference of Dirac Equation from Classical-Mechanics Solution", to be presented at the 5th International Symposium on Quantum Theory and Symmetries, 2007; 
          %%%%%%%%%
(j) (Zheng-Johansson J X) Doebner-Goldin  Equation for electrodynamic model particle. the implied applications {\it Preprint} arxiv:0801.4279
             (talk at  7th  Int.  Conf.  Symm. in Nonl.  Math. Phys., 
               %(Inst. of Math., 
Kyiv, 2007);
              % submitted to  journal for publication; 
(k) Zheng-Johansson J X 2008 
Dirac equation for electrodynamic model particles {\it J. Phys: Conf.  Series} {\bf 128}, 012019, {\it Proc.  5th Int. Symp. Quantum Theory and Symmetries}, ed M Olmo (Valladolid, 2007); 
(l) (Zheng-Johansson J X) 2010 Self interference of single electrodynamic particle in double slit   {\it Proc.  6th Int. Symp. Quantum Theory and Symmetries} ed  A. Shapere and Das (Lexington, 2009) ({Preprint} arxiv:1004.5000);
                  %(J.X. Zheng-Johansson) Quantum probability of IED particle (internal).
                  % preparation (the importance for demonstrating that the IED particle model can consistently represent such systems  was stressed by  Professor H-D Doebner in  recent communications, and by Professor G.A Goldin in discussions at the QTS-4, QTS-5 conferences.) 
(m) (J.X. Zheng-Johansson) 2003 Unification of classical and quantum mechanics \& the theory of relative motion,    {\it Bull.  Amer Phys Soc.} {\bf G35.001}-Gen. Phys. (Austin).
               %(m) (Zheng-Johansson J X 2010 Internally electrodynamic particle model: its experimental basis and its predictions {\it Phys. Atom. Nucl.} {\bf 73}  571-581  ({\it Preprint} arxiv:0812.3951).



%\bibitem{jxzj-Planck} J.X. Zheng-Johansson, Inference of Planck constant and Heisenberg uncertainty relations, 2010  

% \bibitem{Griffith} D Griffith, {\it Introduction to elementary particles}, (1987 Harper and Row Publisher) chapter 7: "Quantum electrodynamics"; J D Bjorken and Drell, S D Drell, {\it Relativistic Quantum Mechanics}, McGraw-Hill cited????

%\bibitem{deBwave-work}We do not see the whole particle  by looking at a bit of its de Broglie wave, and we see the whole only if  we look at it over a  de Broglie wavelength $\lam_d$, for  its minimum energy.





%\bibitem{Niven1981} I Niven, Maxima and Minima without calculus, The Math Assoc Amer, 1981, USA., pp26-28









\end{thebibliography}
\end{document}